\documentclass[prb,aps,twocolumn,epsfig,rotate,showpacs]{revtex4-1}
\usepackage{epsfig}
\usepackage{graphicx}
\usepackage{amsmath}
\usepackage{amssymb}

\newcommand{\beq}{\begin{eqnarray}}
\newcommand{\eeq}{\end{eqnarray}}
\begin{document}
\title{Efficiency of three-terminal thermoelectric transport
\\under broken-time reversal symmetry}
\author{Vinitha Balachandran}
\affiliation{ Center for Nonlinear and Complex Systems,
Universit\`a degli Studi dell'Insubria, Via Valleggio 11, 22100 Como, Italy}
\author{Giuliano Benenti}
\affiliation{CNISM \& Center for Nonlinear and Complex Systems,
Universit\`a degli Studi dell'Insubria, Via Valleggio 11, 22100 Como, Italy}
\affiliation{Istituto Nazionale di Fisica Nucleare, Sezione di Milano,
via Celoria 16, 20133 Milano, Italy}
\author{Giulio Casati}
\affiliation{CNISM \& Center for Nonlinear and Complex Systems,
Universit\`a degli Studi dell'Insubria, Via Valleggio 11, 22100 Como, Italy}
\affiliation{Istituto Nazionale di Fisica Nucleare, Sezione di Milano,
via Celoria 16, 20133 Milano, Italy}
\begin{abstract}
 We investigate thermoelectric efficiency of systems with broken time reversal symmetry under a three-terminal  transport. Using a model of Aharonov-Bohm interferometer formed with three noninteracting quantum dots, we
show that Carnot efficiency $\eta_{C}$ can be achieved when the
thermopower is a symmetric function of the applied magnetic field.
On the other hand, the maximal value of the efficiency at maximum power is obtained for asymmetric thermopower. Indeed, we show that Curzon-Ahlborn limit is exceeded within the linear response regime in our  model. Moreover, we investigate thermoelectric
efficiency for random Hamiltonians drawn from the Gaussian Unitary Ensemble
and for a more abstract transmission model. In this latter model
we find that the efficiency is improved using sharp
energy-dependent transmission functions.

\end{abstract}
\date{\today}
\maketitle

\section{Introduction}
Thermoelectrics convert temperature gradients into electric voltages and vice versa. Strong demand for a cost-effective pollution-free form of energy conversion has resulted in a plethora of thermoelectric-based applications. Unfortunately, the practical usage is limited by extremely low performance of thermoelectric materials.  Increasing the efficiency of thermoelectric materials is one of the main lines of current thermoelectric research \cite{te,Majumdar,dresselhaus,snyder,shakuori,dubi,BC11}.

In the linear response regime, the performance of thermoelectric materials is characterized by a single dimensionless parameter called figure of merit $ZT$,
which is a combination of the main transport properties of a material, i.e.
the electric conductivity $\sigma$, the thermal conductivity
$\kappa$ and the thermopower $S$, as well as of the absolute temperature $T$:
$ZT=(\sigma S^2/\kappa)T$. The maximum efficiency is given by
\begin{equation}
\eta_{max}=\eta_{C}\,\frac{\sqrt{ZT+1}-1}{\sqrt{ZT+1}+1}.
\end{equation}
Carnot efficiency $\eta_{C}$ is reached  in the limit of $ZT \rightarrow \infty$. Although thermodynamics does not impose any upper bound on $ZT$, the strong interdependence between the electric and thermal transport properties makes it extremely hard to increase the value of $ZT$ above $1$. To compete with the existing mechanical based energy conversion applications, development of thermoelectric materials with $ZT$ at least above $3$ is required. An attractive alternative to increase the performance is to use broken time reversal symmetry systems for which the efficiency is determined by two parameters, an asymmetry parameter $x$ and a ``figure of merit'' $y$ generalizing
$ZT$ \cite{prlbts}. In these systems, maximum efficiency $\eta_{max}$ can be as high as Carnot efficiency even with very low values of figure of merit $y$, provided the asymmetry parameter $x$ is very large. Note that in time reversal symmetric systems $x=1$ and $y=ZT$.

In a thermodynamic system, Carnot efficiency is obtained for fully reversible transformations. This requires quasistatic transformations and consequently the derived output power is zero. Hence, the notion of efficiency at maximum power was introduced and in the linear response regime is given by \cite{vandenbroeck}
\begin{equation}
\eta(\omega_{\mathrm{max}})=
\frac{\eta_{C}}{2}\frac{ZT}{ZT+2}.
\end{equation}
Note that in the limit of $ZT \rightarrow \infty$, $\eta(\omega_{\mathrm{max}})$ takes the maximum value of $\eta_{C}/2$. This upper bound is commonly referred to as the Curzon-Ahlborn limit \cite{curzon1,curzon2,curzon3,curzon4,vandenbroeck,esposito2009,schulman,esposito2010,linke,seifert,goupil}. In principle, for broken time reversal symmetry systems this limit can be exceeded with values of the asymmetry parameter such that $|x|>1$ \cite{prlbts}. Moreover, $\eta(\omega_{\mathrm{max}})\rightarrow \eta_{C}$ when $|x|\to\infty$ (always within the linear response regime).  Hence, it is potentially of practical relevance to study the thermoelectric transport in broken time reversal symmetry systems.

Thermopower for a broken time reversal symmetry system is in general asymmetric with respect to the time reversibility breaking parameter $\mathbf{B}$ i.e., $S(\mathbf{B})\neq S(-\mathbf{B})$.
The asymmetry parameter $x=S(\mathbf{B})/ S(-\mathbf{B})$. In a non-interacting system, inelastic scattering can result in asymmetric thermopower \cite{astp1,astp2}. Conveniently, this can be achieved with the introduction of noise by means of a third terminal (probe) with its temperature and chemical potential adjusted such that there is no net average flux of particles and heat between the terminal
and the system.
Various aspects of three-terminal thermoelectric transport were
investigated in Refs. \cite{imry1,buttiker,imry2,ora,buttiker2}
Large asymmetry in the thermopower were obtained in Ref.~\cite{astp1},
using an Aharonov-Bohm interferometer model, first treated in the context
of thermoelectric transport in Ref.~\cite{ora}.
However, the obtained efficiency was very low \cite{astp1}. Also, following the above lines, a classical deterministic three terminal transport model was studied, showing large asymmetry with very low efficiency \cite{railway}. Hence, it remains to be seen whether and under what conditions efficiency close to Carnot can be obtained using broken time reversal symmetry system under a three terminal transport.

In this paper, we investigate the thermoelectric efficiency of non-interacting systems with broken time reversal symmetry under three terminal transport in the linear response regime. First we consider a model of Aharonov-Bohm interferometer formed with three dots as analyzed in Ref. \cite{astp1}.  In contrast to the earlier work, we focus on the efficiency obtained over the global optimization of all system and reservoir parameters following simulated annealing method \cite{sti}. Our results show that Carnot efficiency $\eta_{C}$ can be obtained for maximum efficiency $\eta_{max}$ when the thermopower is symmetric i.e., $x=1$.  By adding asymmetry, $\eta_{max}$ decreases. However, the efficiency at maximum power $\eta(\omega_{\mathrm{max}})$ is maximum when there is asymmetry in thermopower ($x\neq1$).
To the best of our knowledge for the first time we show that the
Curzon-Ahlborn limit $\eta(\omega_{\mathrm{max}})=\eta_{C}/2$, which is
a rigorous upper bound for systems with time-reversal symmetry,
is exceeded in the linear response regime.
However, both the efficiency $\eta_{max}$ and  $\eta(\omega_{\mathrm{max}})$  decrease with increase in asymmetry of thermopower at large values of $|x|$.
Therefore, we study broader classes of models in an attempt to
improve efficiency at large asymmetry.
We consider random Hamiltonians drawn from Gaussian Unitary Ensemble (GUE)
and finally an abstract model of transmission probabilities.
While in this latter model we show that $\eta(\omega_{\mathrm{max}})$ can take values as high as $0.57\eta_{C}$, still we could not find, after global
optimization of all parameters of the model, large efficiency at large
asymmetry.
Our results obtained for very broad classes of non-interacting models
implies that it is
practically very hard, if not impossible, to achieve in such models
large efficiency with large asymmetry in thermopower.

The paper is structured as follows: In Sec. II, we review the calculations of efficiency of broken time reversal symmetry systems under three terminal transport. Dependence of efficiency on asymmetry of thermopower for various models with broken time reversal symmetry is analyzed in Sec. III. Finally, Sec. IV summarizes our results.

\section{Model and Method}

\subsection{General setup}
The general set up  consists of a system in contact with two reservoirs left (L) and right (R) at temperatures $T_{L}=T+\Delta T,T_{R}=T$ and chemical potentials $\mu_{L}=\mu+\Delta \mu,\mu_{R}=\mu$. Inelastic scattering effects are simulated by means of a third (probe) reservoir (P) at temperature $T_{P}=T+\Delta T_{P}$ and chemical potential $\mu_{P}=\mu+\Delta \mu_{P}$. Let $J_{\rho k}$ and $J_{E k}$ denote the electric and energy currents from the $k$th reservoir ($k$=L,R,P) into the system, with the steady-state constraints of charge and energy conservation $\sum_{k} J_{\rho k} =0,\sum_{k} J_{E k} =0$. The sum of the entropy production rates at the reservoirs reads  $\dot{S}=\sum_{k}(J_{E k}-\mu_{k} J_{\rho k})/ T_{k}$.  Within linear response, $\dot{S}=\mathbf{J}\cdot \mathbf{X}\equiv\sum_{i=1}^{4} J_{i}X_{i}$, where $\mathbf{J}$ and $\mathbf{X}$ are four dimensional vectors defined as
\begin{eqnarray}
  \mathbf{J} &=& (e J_{\rho L} , J_{q L}, e J_{\rho P}, J_{q P}),\\
  \mathbf{X} &=& (\frac{\Delta \mu}{e T},\frac{\Delta T}{ T^{2}},\frac{\Delta\mu_{P}}{e T},\frac{\Delta T_{P}}{ T^{2}}).
\end{eqnarray}
Here $e$ is the electron charge and $J_{qk}\equiv J_{E k}-\mu  J_{\rho k}$ is the heat current.
The relation between the fluxes $J_{i}$ and the thermodynamic forces $X_{i}$ within linear irreversible thermodynamics is
\begin{equation}\label{flux}
    \mathbf{J}=\mathbf{L}\mathbf{X},
\end{equation}
where $\mathbf{L}$ is a $4\times4$ Onsager matrix, $\mathbf{J}$ and $\mathbf{X}$ are written as column vectors. Eq. (\ref{flux}) can be written in the block matrix form as
\begin{eqnarray}\label{mut}
 \left( \begin{array}{c}
 \mathbf{ J}_{\alpha} \\
 \mathbf{ J}_{\beta}
\end{array}
\right) &=&  \left( \begin{array}{cc}
             \mathbf{ L}_{\alpha\alpha} &  \mathbf{L}_{\alpha\beta} \\
              \mathbf{L}_{\beta\alpha} & \mathbf{L}_{\beta\beta}
            \end{array} \right) \left( \begin{array}{cc}
             \mathbf{ X}_{\alpha} \\
             \mathbf{X}_{\beta}
            \end{array} \right),
\end{eqnarray}
where $\alpha$ stands for $(1,2)$ and $\beta$ for $(3,4)$.

The probe reservoir is adjusted in such a way that $\mathbf{J}_{\beta}=J_{3}=J_{4}=0$, that is, the net electric and heat flow from the probe into the system vanishes. This implies that $\mathbf{X}_{\beta}=-\mathbf{L}_{\beta\beta}^{-1}\mathbf{L}_{\beta\alpha}\mathbf{X}_{\alpha}$ and
\begin{equation}\label{flux2}
   \mathbf{J}_{\alpha}=\mathbf{L}'\mathbf{X}_{\alpha}, \quad \mathbf{L}'\equiv \mathbf{L}_{\alpha\alpha}-\mathbf{L}_{\alpha\beta}\mathbf{L}^{-1}_{\beta\beta}\mathbf{L}_{\beta\alpha}.
\end{equation}
Thus, the problem has been reduced to two coupled fluxes as
\begin{eqnarray}\label{mut1}
 \left( \begin{array}{c}
  J_{1} \\
  J_{2}
\end{array}
\right) &=&  \left( \begin{array}{cc}
              L'_{11} &  L'_{12} \\
              L'_{21} & L'_{22}
            \end{array} \right) \left( \begin{array}{cc}
              X_{1} \\
             X_{2}
            \end{array} \right),
\end{eqnarray}
where the reduced $2\times2$ Onsager matrix $\mathbf{L}'$ satisfies the Onsager-Casimir relations
\begin{equation}\label{casimir}
    L'_{ij}(\mathbf{B})=L'_{ji}(-\mathbf{B}) \quad(i,j=1,2).
\end{equation}
Here, $\mathbf{B}$ is a magnetic field breaking the time reversal symmetry. Seebeck and Peltier coefficients are given by  $S=L_{12}'/(eTL_{11}')$ and $\Pi=L_{21}'/(eL_{11}')$. Thermopower is asymmetric when $L_{12}'\neq L_{21}'$, i.e., $\Pi\neq ST$.

Maximum efficiency is given by
\begin{equation}\label{powermax}
    \eta_{max}=\eta_{C}x\frac{\sqrt{y+1}-1}{\sqrt{y+1}+1} ,
\end{equation}
 and depends on two parameters: the asymmetry parameter $x$ and
the figure of merit $y$, where
 \begin{eqnarray}\label{as}
  x &\equiv& \frac{L'_{12}}{L'_{21}} ,\\
  y &\equiv& \frac{L'_{12}L'_{21}}{\det \mathbf{L}'}.
\end{eqnarray}
Efficiency at maximum power is
\begin{equation}\label{effmax2}
    \eta(\omega_{max})=\frac{\eta_{C}}{2}\frac{xy}{2+y}.
\end{equation}
Although the thermodynamics does not impose any restriction on the
attainable values of asymmetry parameter $x$, the positivity of
entropy production rate implies that
\begin{align}
&h(x)\leq y \leq0& &\mathrm{if} \quad x<0, \nonumber \\
&0 \leq y \leq h(x)& &\mathrm{if} \quad x>0,
              \end{align}
where $h(x)=4x/(x-1)^{2}$. Maximum values
of both $\eta_{max}$ and
$\eta(\omega_{max})$ are obtained, for a given x, when $y=h(x)$.
We denote such maximum values as
$\eta_{max}^\star$ and
$\eta(\omega_{max})^\star$, respectively.
From Eqs.
(\ref{powermax}) and (\ref{effmax2}), it follows that the
theoretical upper bounds are given, for maximum efficiency, by
\begin{align}
  \eta_{max}^{\star}&=\eta_{C}x^{2}& &\mathrm{if} \quad |x|<1, \nonumber \\
   \eta_{max}^{\star}&=\eta_{C}&     &\mathrm{if} \quad |x|\geq1,
\label{etamaxstar}
\end{align}
and, for efficiency at maximum power, by
\begin{equation}\label{bound2}
   \eta(\omega_{max})^\star=\eta_{C}\frac{x^{2}}{x^{2}+1}.
\end{equation}
It is clear from the above equation that Curzon-Ahlborn limit can
in principle be overcome for broken-time reversal symmetry systems with
asymmetry parameter $|x|>1$ \cite{prlbts}.

\subsection{Noninteracting systems}
Thermoelectric efficiency can be calculated exactly for
noninteracting models by means of Landauer-B\"uttiker formalism
\cite{Landauer}. Consider then a noninteracting system with
Hamiltonian $H_{S}$.  We model the reservoirs as ideal Fermi gases
with Hamiltonian
\begin{equation}\label{res}
H_{R}=\sum_{qk}E_{qk}d_{qk}^{\dagger}d_{qk}.
\end{equation}
Here, $E_{qk}$ is the energy of an electron in the
state $q$ in the $k$th
reservoir, and $d_{qk}^{\dagger}$ and $d_{qk}$ are the
corresponding creation and
annihilation operators.

 The
electric and heat currents from the left reservoir are given by
\begin{eqnarray}\label{current}
  J_{1} &=& \frac{e}{h}\int dE \sum_{k}[\tau_{kL}(E)f_{L}(E)-\tau_{Lk}(E)f_{k}(E)],\nonumber\\
   J_{2} &=& \frac{1}{h} \int dE \sum_{k} (E-\mu_{L}) [\tau_{kL}(E)f_{L}(E)\nonumber\\
   &&\qquad -\tau_{Lk}(E)f_{k}(E)],
\end{eqnarray}
where $\tau_{kl}(E)$ is the transmission probability from reservoir
$l$ to reservoir $k$ at energy $E$ and
$f_{k}(E)=\{\mathrm{exp}[(E-\mu_{k})/k_{B}T_{k}]+1\}^{-1}$ is the
Fermi distribution function. Analogous expressions can be written
for $J_{3}$ and $J_{4}$, provided the terminal $L$ is substituted by
$P$.

The Onsager coefficients $L_{ij}$ are obtained from the linear response expansion of the currents $J_{i}$ as
\begin{eqnarray}\label{onsager}
   L_{11} &=& \frac{e^2T}{h}\int_{-\infty}^{+\infty} dE \,\sum_{k\neq L}\tau_{Lk}(E)
\left[-\frac{\partial f(E)}{\partial E}\right], \nonumber\\
  L_{12} &=& \frac{eT}{h}\int_{-\infty}^{+\infty} dE \, (E-\mu) \sum_{k\neq L}\tau_{Lk}(E)
\left[-\frac{\partial f(E)}{\partial E}\right], \nonumber\\
  L_{22} &=& \frac{T}{h}\int_{-\infty}^{+\infty} dE \, (E-\mu)^{2} \sum_{k\neq L}\tau_{Lk}(E)
\left[-\frac{\partial f(E)}{\partial E}\right], \nonumber\\
  L_{21} &=& L_{12}.
\label{eq:Lij}
\end{eqnarray}
Analogous formulas are obtained for $L_{33}$, $L_{34}=L_{43}$, and $L_{44}$, with the $P$ terminal used instead of $L$. Similarly, the off diagonal block elements are obtained as
\begin{eqnarray}\label{onsager1}
   L_{13} &=& -\frac{e^2T}{h}\int_{-\infty}^{+\infty} dE \,\tau_{LP}(E)
\left[-\frac{\partial f(E)}{\partial E}\right], \nonumber\\
  L_{14} &=& -\frac{eT}{h}\int_{-\infty}^{+\infty} dE \, (E-\mu) \tau_{LP}(E)
\left[-\frac{\partial f(E)}{\partial E}\right], \nonumber\\
  L_{24} &=& -\frac{T}{h}\int_{-\infty}^{+\infty} dE \, (E-\mu)^{2} \tau_{LP}(E)
\left[-\frac{\partial f(E)}{\partial E}\right], \nonumber\\
  L_{23} &=& L_{14}.
\label{eq:Lij1}
\end{eqnarray}
Using $\tau_{PL}(E)$ instead of $\tau_{LP}(E)$ in Eq. (\ref{onsager1}), $L_{31}$, $L_{32}=L_{41}$, and $L_{42}$ are obtained.

The transmission probabilities $\tau_{pq}(E)$ are calculated as
\begin{equation}\label{tp}
    \tau_{pq}(E)={\rm Tr}[\Gamma_{p}(E)G(E)\Gamma_{q}(E)G^{\dag}(E)],
\end{equation}
 where the broadening matrices $\Gamma_{k}$ are defined in terms of the self-energies $\Sigma_{k}$ as $\Gamma_{k} \equiv i [\Sigma_{k}(E)-\Sigma_{k}^{\dag}(E)]$ and the (retarded) system Green function $G(E)\equiv[E-H_{S}-\sum_{k} \Sigma_{k}(E)]^{-1}$.

Explicit expression for the matrix elements of the reduced  $2\times2$
Onsager matrix $\mathbf{L}'$ can be derived from
Eq.~(\ref{flux2}). We obtain
\begin{eqnarray}\label{int1}
  L'_{11} &=& L_{11}-\frac{e^2}{D}  \int_{\mathbb{R}^3} dI (E_{2}-E_{1}) (E_{2}-E_{3}), \nonumber \\
   L'_{22} &=& L_{22}-\frac{1}{D}  \int_{\mathbb{R}^3} dI (E_{3}-\mu)(E_{1}-\mu) (E_{2}-E_{1})(E_{2}-E_{3}), \nonumber \\
    L'_{12} &=& L_{12}-\frac{e}{D}  \int_{\mathbb{R}^3} dI (E_{3}-\mu)(E_{2}-E_{1}) (E_{2}-E_{3}), \nonumber \\
   L'_{21} &=& L_{21}-\frac{e}{D}  \int_{\mathbb{R}^3} dI (E_{1}-\mu) (E_{2}-E_{1}) (E_{2}-E_{3}).
\end{eqnarray}
  Here $dI$ and $D$ are
\begin{eqnarray}\label{int}
   dI &=& \frac{e^2T^3}{h^3} \left[\prod_{i=1}^{3} dE_{i} \left(-\frac{\partial f}{\partial E_{i}}\right)\right]  \nonumber \\ &&\times \tau_{LP}(E_{1}) [\tau_{PL}(E_{2})+ \tau_{PR}(E_{2})]\tau_{PL}(E_{3}), \nonumber \\
   D &=& {\rm det}\left(\mathbf{L}_{\beta\beta}\right)
=\left(\frac{eT}{h}\right)^2
\int_{\mathbb{R}^2}\left\{\prod_{i=1}^{2} dE_{i} \left(-\frac{\partial f}{\partial E_{i}}\right)\right.  \nonumber \\ &&\left.[\tau_{PL}(E_{i})+\tau_{PR}(E_{i})]\right\}  (E_{1}-\mu) (E_{1}-E_{2}).
\end{eqnarray}


\section{Results}
\subsection{Aharonov-Bohm interferometer}

\begin{figure}
  \begin{center}
 \epsfig{file=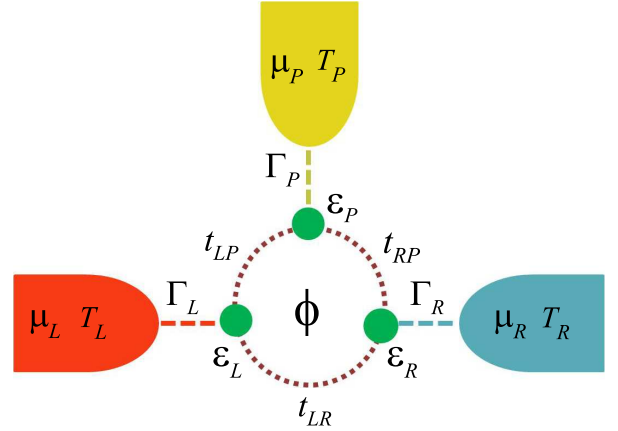,width=9cm}
  \end{center}
  \caption{Schematic picture of
Aharonov-Bohm interferometer model.} \label{sch}
\end{figure}

Here, we discuss the thermoelectric efficiency of a three quantum
dot ring structure pierced by an Aharonov-Bohm flux and coupled to
three reservoirs, with each dot connected independently to one
reservoir. Figure \ref{sch} displays a sketch of the model. The system is described by the Hamiltonian
\begin{eqnarray}\label{abhamil}
    H_{S}&=&\sum_{k}\epsilon_{k}c_{k}^{\dag}c_{k}+(t_{LR}c_{R}^{\dag}c_{L}e^{i\phi/3} \nonumber\\
    &+&t_{RP}c_{P}^{\dag}c_{R}e^{i\phi/3}
+t_{PL}c_{L}^{\dag}c_{P}e^{i\phi/3}+ \hbox{H.c.}),
\end{eqnarray}
where $\epsilon_{k}$ are the on-site energies, $t_{ij}$ are the
hopping strengths $(i,j=L,R,P)$, $\phi$ is the flux.
$c_{k}^{\dagger}$ ($c_{k}$) are the creation (annihilation)
operators of the electron in the $k$th dot. Reservoirs are ideal fermi gases with Hamiltonian given by Eq. (\ref{res}). The dot-reservoir
coupling Hamiltonian is
\begin{eqnarray}\label{coupl}
    H_{SR}&=&\sum_{q}(t_{qL}c_{L}^{\dag}d_{qL}+
    t_{qR}c_{R}^{\dag}d_{qR}\nonumber\\ &+& t_{qP}c_{P}^{\dag}d_{qP} +\hbox{H.c.}).
\end{eqnarray}
Here, $t_{qj}$ is the tunneling amplitude of an electron in the
state $q$ into the
$k$th reservoir. We assume the wide-band limit and hence the
broadening matrices are given by
$\Gamma_{k}=\gamma_{k}c_{k}^{\dag}c_{k}$, where $\gamma_{k}=
2 \pi \sum_{q}|t_{qk}|^{2}\delta(E-E_{qk})$.  Note that $\gamma_{k}$
measures the tunneling rate of electrons between the reservoir $k$
and the system.

We follow  Landauer-B\"uttiker formalism and calculate Onsager coefficients from Eqs.(\ref{onsager}) to (\ref{tp}).
When there is  anisotropy in the system $(\epsilon_{k} \neq \epsilon_{j})$,
 and the Aharonov-Bohm flux $\phi$ is non-zero, the off diagonal elements of the reduced Onsager matrix $\mathbf{L}'$ are asymmetric functions of the
flux, i.e., $L'_{12}(\phi)\neq L'_{21}(\phi)=L'_{12}(-\phi)$.
The asymmetry parameter $x$ defined in Eq. (\ref{as}) is the ratio of the off diagonal Onsager matrix elements and is in general
different from the time reversal symmetric case where $x=1$.

\begin{figure}
  \begin{center}
 \epsfig{file=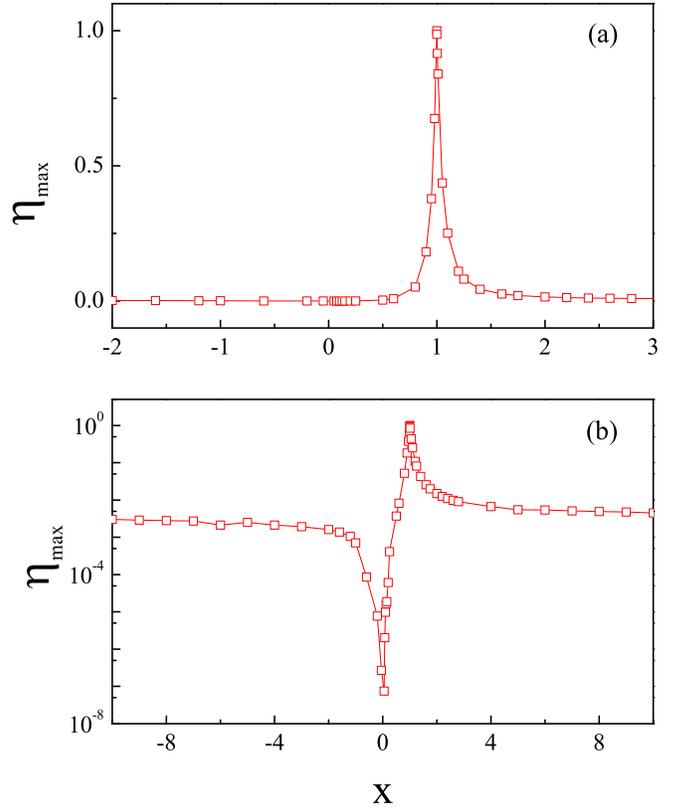,width=9cm}
  \end{center}
  \caption{Optimized values of maximum efficiency $\eta_{max}$ as a function of the asymmetry parameter $x$ for the
Aharonov-Bohm interferometer discussed in the text, with a linear
[panel (a)] and a logarithmic [panel (b)] scale for $\eta_{\rm max}$. Note that $\eta_{max}$ is scaled in units of $\eta_{C}$ in this figure and thereafter.   } \label{eff}
\end{figure}

For this model, the efficiency is a function of $12$ independent parameters: $\epsilon_{1}$, $\epsilon_{2}$, $\epsilon_{3}$, $t_{LR}$, $t_{RP}$, $t_{PL}$, $\phi$,
$\gamma_{L}$, $\gamma_{R}$, $\gamma_{P}$, $T$, $\mu$. We maximize the efficiency over all these parameters using the simulated annealing method. For numerical convenience, the parameters are restricted to the intervals
$\epsilon_{i}\in [-5,5]$, $t_{ij}\in[-5,5]$, $T\in [10^{-3},100]$, $\mu\in [-5,5]$, $\phi\in [-1,1]$, $\gamma_{i}\in [10^{-3},1]$
(in units such that $e=h=k_{B}=1$).
Note that the results are unchanged by varying the range of
parameter values by a few times.
Fig. \ref{eff}(a) shows the dependence of the
numerically obtained
maximum efficiency $\eta_{max}$ on the asymmetry parameter $x$
\cite{footnote:cost}.
In the absence of asymmetry in the thermopower i.e., when $x=1$, Carnot efficiency  is reached. As the asymmetry is introduced, the optimized maximum efficiency is always less than Carnot efficiency.

\begin{figure}[h!]
  \begin{center}
 \epsfig{file=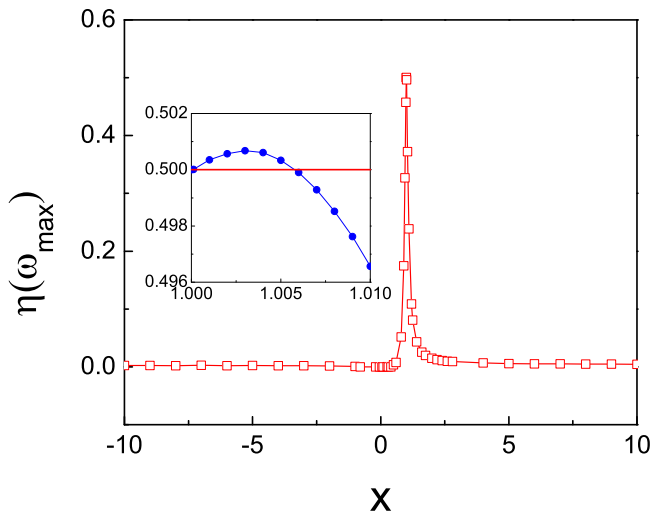,width=10cm}
  \end{center}
  \caption{Optimized values of efficiency at maximum power $\eta(\omega_{max})$ versus asymmetry parameter $x$ for the Aharonov-Bohm interferometer. The inset shows the variation of $\eta(\omega_{max})$ near the symmetric value $x=1$. Note that the Curzon-Ahlborn limit $\eta(\omega_{max})=\eta_{C}/2$  is exceeded in this regime. $\eta(\omega_{max})$ is scaled in units of $\eta_{C}$ in this figure and thereafter. } \label{effc}
\end{figure}

We also plot the dependence of the optimized maximum efficiency $\eta_{max}$ on $x$ in a lin-log scale in Fig. \ref{eff}(b). It is clear from the figure that for $|x|<1$,  $\eta_{max}$ increases with $|x|$ from its zero value at $x=0$. Also, the increase is more drastic with positive values of $x$.  Indeed, only for $x>0$ Carnot efficiency is reached. For negative values of $x$, $\eta_{max}$ increases initially
(up to $x=-10$ where $\eta_{\rm max}=0.003$) and decreases thereafter.

Similar results were obtained by maximizing the efficiency at maximum power $\eta(\omega_{max})$. This is illustrated in Fig. \ref{effc}. For $x=1$, $\eta(\omega_{max})=\eta_{C}/2$  and the Curzon-Ahlborn limit is recovered.  From our results we find that at large asymmetry,  $\eta(\omega_{max})$ decreases with the increase of $|x|$.
It is interesting to remark that,
in contrast to $\eta_{max}$, the maximum value of $\eta(\omega_{max})$ is obtained when $x\neq1$. In particular, we find that very near to the symmetric value $x=1$, the Curzon-Ahlborn limit of efficiency at maximum power $\eta(\omega_{max})=\eta_{C}/2$  can be exceeded, as shown in the inset of Fig. \ref{effc}. Note that in Landauer-B\"uttiker formalism, the obtained results are exact. Hence our results are bound only by machine accuracy and the calculated efficiency is accurate up to $15$ decimal points.
Even though there is only a small improvement with respect to the
Curzon-Ahlborn limit, such result is interesting in that it provides
the first evidence, in a concrete model, of the fact that the Curzon-Ahlborn
limit, which is a universal upper bound for time-reversal systems within
linear response, can be exceeded when time reversibility is broken.

According to our numerical optimization, large asymmetry results in low
thermoelectric efficiency.
Also, we did not observe a significant improvement in efficiency
by increasing the number of levels in the system up to six
(for systems with a larger number of levels, it becomes difficult to
obtain convergence).

\subsection{Gaussian Unitary Ensemble model}
\begin{figure}[h!]
  \begin{center}
  \epsfig{file=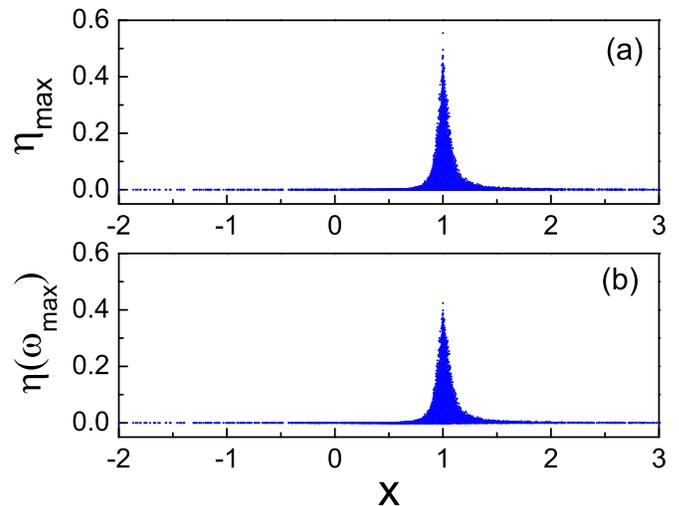,width=9cm}
  \end{center}
  \caption{Relation between maximum efficiency $\eta_{max}$ and asymmetry parameter $x$
[panel (a)] and between efficiency at maximum power $\eta(\omega_{max})$ and asymmetry parameter
[panel (b)]
for $20000$ realizations of  $3\times3$ Hamiltonian drawn from
the Gaussian Unitary Ensemble.} \label{gue}
\end{figure}

To see whether thermoelectric efficiency can be improved in more general systems, we turn to Random Matrix Theory (RMT) models. Indeed, Hamiltonians drawn from Gaussian Unitary Ensemble (GUE) give a good description of complex physical systems with broken time reversal symmetry.  To this end, in this section we study the efficiency
(and asymmetry) distribution of GUE Hamiltonians.

We consider a random Hamiltonian $H_{S}$  drawn from a
Gaussian Unitary Ensemble \cite{haake}.  That is,
matrix elements $H_{lm}$ of the Hamiltonian $H_S$ are chosen such that the diagonal entries are normally distributed with zero mean and unit variance ($N(0,1)$) and the off-diagonal entries are drawn independently and identically from the normal distribution (subject to being Hermitian) with mean zero and variance $\frac{1}{2}$ $(N(0,\frac{1}{2}))$ . In other words, $H_{lm}=U_{lm}+ i V_{lm}$, where $U_{lm},V_{lm}\in N(0,\frac{1}{2})$, for  $1\leq l < m \leq n$, $H_{ml}=H_{lm}^\star$,
and $H_{mm} \in N(0,1)$ for $1\leq m\leq n$,
with $n$ the dimension of the matrix.

To compare with the results of the Aharonov-Bohm interferometer, we set the dimension of the matrix as $n=3$. Also, dot-reservoir coupling is taken to be same as that of the Aharonov-Bohm interferometer discussed in the previous subsection. Results of one such calculation with $20000$ different realizations are plotted in Fig. \ref{gue}. Here, we take  $T=0.1$, $\mu=0.1$ and $\gamma_{L}=\gamma_{R}=\gamma_{P}=0.1$. Note that results are not sensitive to variations of the parameters $T$, $\mu$, $\gamma_{L}$, $\gamma_{R}$,
$\gamma_{P}$. Top panel represents the dependence of maximum efficiency $\eta_{max}$ on the asymmetry parameter $x$  and bottom panel the dependence of efficiency at maximum power $\eta(\omega_{max})$ on $x$. Results shown in the figure reveal that as the asymmetry in the thermopower is increased, the efficiency decreases. This implies that it is practically hard to achieve  large efficiency with large asymmetry in thermopower and confirms our results obtained by optimizing the efficiency of Aharonov-Bohm interferometer in the previous subsection.

\begin{figure}
  \begin{center}
  \epsfig{file=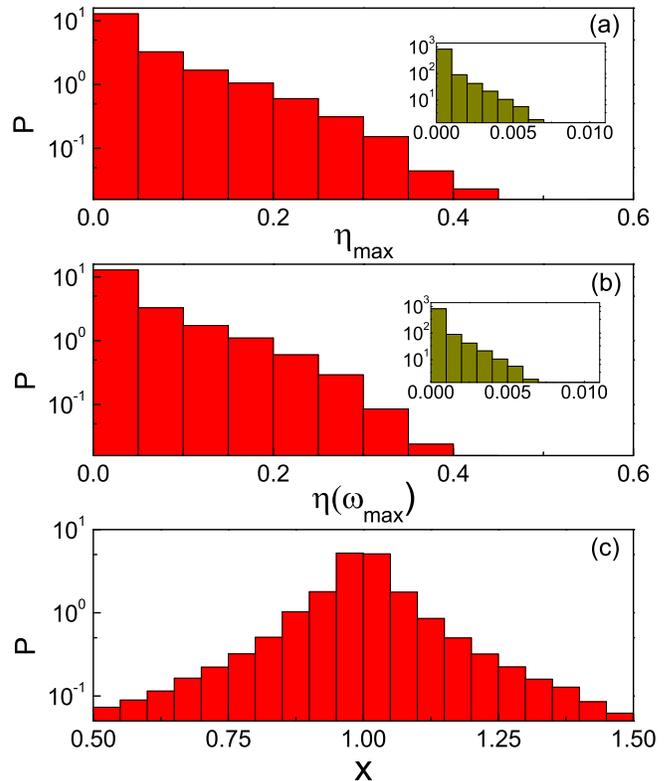,width=9.cm}
  \end{center}
  \caption{Distribution of (a) maximum efficiency  $\eta_{max}$, (b) efficiency at maximum power $\eta(\omega_{max})$ and (c) asymmetry parameter $x$. Inset of (a) and (b) shows the distribution of   $\eta_{max}$ and $\eta(\omega_{max})$, taking into account only data such that $|x-1|>0.5$.  Note that obtaining high values of both efficiency and asymmetry is highly improbable.  } \label{stat}
\end{figure}

Actually, we find that obtaining large values of efficiency is highly improbable.
This is illustrated in Fig. \ref{stat} with the logarithmic distribution of  $\eta_{max}$ and $\eta(\omega_{max})$. Both $\eta_{max}$ and $\eta(\omega_{max})$  follows an exponentially decaying distribution indicating that large efficiency is exponentially improbable.
Also, the distribution of asymmetry parameter $x$ follows exponential distribution implying that large asymmetry is also a rare situation in this model.
Note that the maximum value of efficiency is obtained around the symmetric point $x=1$. Hence, to better clarify the role of asymmetry we have plotted the logarithmic distribution of $\eta_{max}$ and $\eta(\omega_{max})$, restricted to data for which $|x-1|>0.5$, in the inset of Fig. \ref{stat}(a) and  \ref{stat}(b) respectively. From the figure, it is clear that distribution of $\eta_{max}$ and $\eta(\omega_{max})$  is exponential even in the absence of symmetry.
On the other hand, there is no numerical evidence of a cut-off or a sharp border
forbidding $\eta=\eta_C$ at $|x|>1$.

In order to study the dependence of efficiency on the number of levels of the system, we studied GUE Hamiltonians by increasing the dimension $n$. Our results with $n=4$ and $n=5$ (data not presented) show that there is little improvement in efficiency with the increase in the levels of the system.

Thus, having studied both the Aharonov-Bohm interferometer model and
the broader class of GUE Hamiltonians,
we can conclude that for Hamiltonian models it is difficult to obtain large values of efficiency with large asymmetry. Hence, in the next subsection we turn to a more abstract transmission model and investigate whether the efficiency can be improved.

\subsection{Transmission model}

Here, we assume that the information about the system is inaccessible and only the scattering matrix that defines the transmission probabilities $\tau_{ij}(E)$ at each energy $E$ are known. Onsager matrix elements are calculated directly from Eqs. (\ref{onsager}) and (\ref{onsager1}) by substituting the values of transmission probabilities $\tau_{ij}(E)$. These transmission probabilities are bound to follow $\sum_{i}\tau_{ij}(E)=1$ that implies the conservation of probability, and the sum rule $\sum_{i}\tau_{ij}(E)=\sum_{j}\tau_{ij}(E)$ that ensures
that the currents vanish at equilibrium.
To simplify the calculations, the transmission probabilities are taken to be constant over a window of energy.
More precisely,
each energy window is characterized by two parameters: $\bar{E}_{k}$, the center of the window and  $\Delta_{k}$, the width of the window. Thus, in our model $\tau_{ij}(E)=\left(c_{ij}\right)_k$
($0\le \left(c_{ij}\right)_k \le 1$) when energy $E\in[\bar{E}_{k}-\Delta_{k}/2,
\bar{E}_k+\Delta_k/2]$, and $\tau_{ij}(E)=0$ otherwise.
Efficiency is optimized over the parameters temperature $T$, chemical potential $\mu$, energies $\bar{E}_{k}$, width $\Delta_k$,
and transmission probabilities $\left(c_{ij}\right)_k$.
We consider $n$ non-overlapping transmission windows
($k=1,...,n$).

First, we consider a single energy window of transmission ($n=1$).
Here, the integration domain $\mathbb{R}^3$ in Eq. (\ref{int1}) is restricted to
the cube $\bar{E}_1-\Delta_1/2\le E_i\le
\bar{E}_1+\Delta_1/2$ ($i=1,2,3$) and
\begin{eqnarray}
L'_{12}-L'_{21}&=&
A \int_{\mathbb{R}^3}
\left[\prod_{i=1}^{3} dE_{i}
\left(-\frac{\partial f}{\partial E_{i}}\right)\right]
\nonumber\\
&\times&
(E_3-E_1)(E_2-E_1)(E_2-E_3),
\label{intsym}
\end{eqnarray}
where the constant
\begin{equation}
A=\frac{e^3T^3}{h^3 D} \left(c_{LP}\right)_1
\left[\left(c_{PL}\right)_1+\left(c_{PR}\right)_1\right]
\left(c_{PL}\right)_1.
\end{equation}
The integral in Eq. (\ref{intsym}) vanishes since it is an
odd function of, for instance, $E_1-E_3$ and the integration domain is
symmetric under exchange of $E_1$ and $E_3$.
Thus, $L'_{12}= L'_{21}$  and  a symmetric thermopower is obtained.

For $n=2$ energy windows, we found numerically that
the asymmetry parameter $x$ is always limited to a finite interval.
Infinitely large values of asymmetry are obtained for transmission with at least three energy windows. For $n=3$ windows, we need to optimize over $20$ parameters:
 $T$, $\mu$, $\bar{E}_{k}$, $\Delta_{k}$, and $\left(c_{LR}\right)_k$,
$\left(c_{LP}\right)_k$,
$\left(c_{PL}\right)_k$,
$\left(c_{PR}\right)_k$,
($k=1,2,3$). (The other transmissions $\left(c_{ij}\right)_k$ are
then determined from the conditions $\sum_i \left(c_{ij}\right)_k=
\sum_j \left(c_{ij}\right)_k=1$.)
\begin{figure}[h!]
  \begin{center}
  \epsfig{file=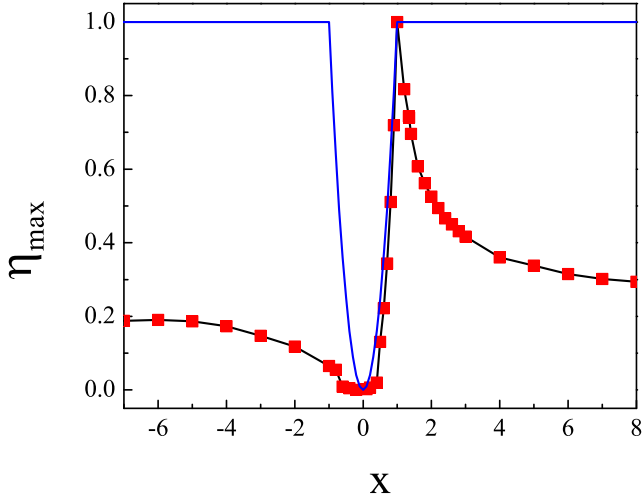,width=10cm}
  \end{center}
  \caption{Maximum efficiency $\eta_{max}$  versus asymmetry parameter $x$ under three terminal transport with transmission model described in the text. Solid curve corresponds to the theoretically predicted upper bound
$\eta_{\rm max}^\star$ (given by Eq.~(\ref{etamaxstar}))
for  $\eta_{max}$   for a generic broken time reversal symmetry system. The optimized efficiency
saturates the theoretical upper bound
 only near the symmetric value $x=1$.  } \label{trans}
\end{figure}

Fig. \ref{trans} shows the variation of the optimized maximum efficiency $\eta_{max}$ with the asymmetry parameter $x$. As in the case of the Aharonov-Bohm interferometer,  Carnot efficiency $\eta_{C}$ is obtained only in the symmetric case $x=1$. Introducing asymmetry results in the reduction of maximum efficiency from $\eta_{C}$. In particular, for positive values of asymmetry, $\eta_{max}$ increases quadratically from zero value to $\eta_{C}$ at $x=1$ and decreases thereafter.  When the asymmetry parameter is negative,  $\eta_{max}$  increases
with $-x$ from $x=0$ to $x=-50$ (where $\eta_{max}=0.24262$) and then
decreases for larger values of $-x$.

\begin{figure}
  \begin{center}
  \epsfig{file=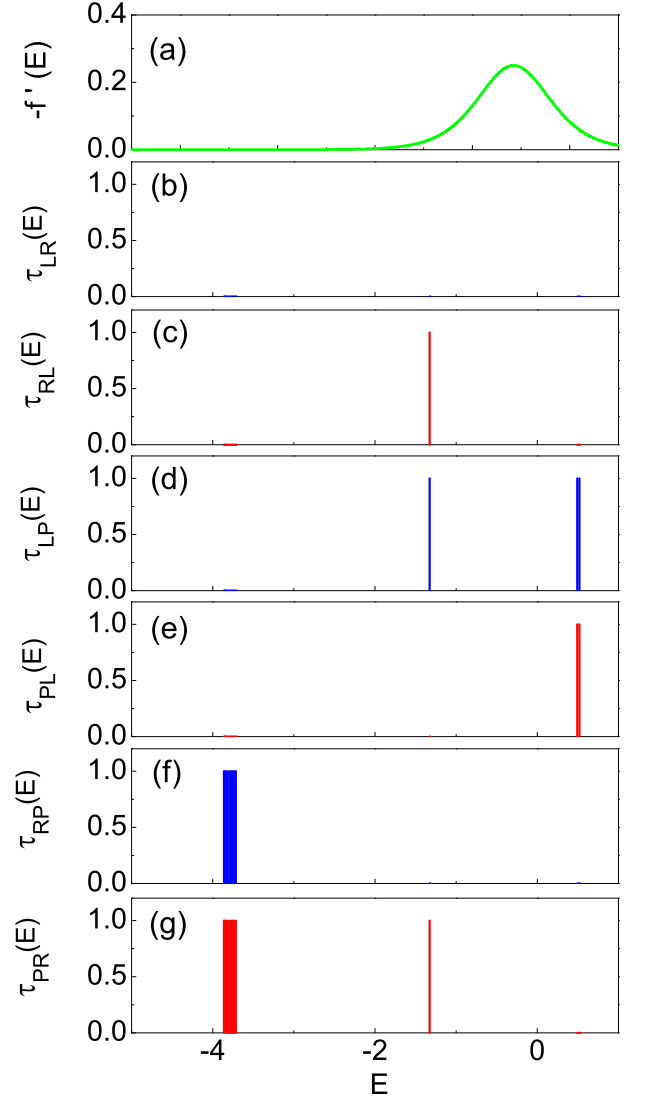,width=10cm}
  \end{center}
  \caption{Optimized transmission functions for $x=1.4$ using the transmission model. Panel (a) shows $-f'$, with $f'$ derivative of the Fermi distribution function $(f'(E)=\frac{\partial f(E)}{\partial E})$, as a function of energy $E$ with optimal temperature $T=1$ and chemical potential $\mu=1.3333$.  The transmission probabilities $\tau_{LR}$, $\tau_{RL}$, $\tau_{LP}$, $\tau_{PL}$, $\tau_{RP}$ and $\tau_{PR}$  at each energy $E$ are plotted in panels (b), (c), (d), (e), (f) and (g) respectively. Note that optimal transmission probabilities are either $0$ or $1$. } \label{opt}
\end{figure}

\begin{figure}[h!]
  \begin{center}
  \epsfig{file=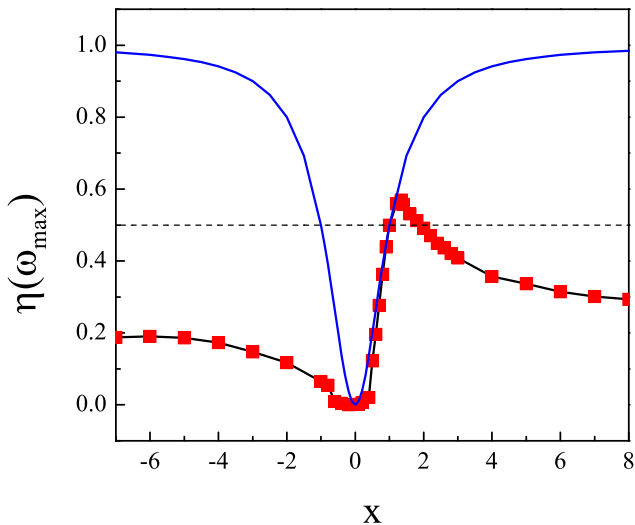,width=10cm}
  \end{center}
  \caption{Efficiency at maximum power $\eta(\omega_{max})$  versus asymmetry parameter $x$ under three terminal transport using the transmission model described in the text. Theoretically predicted upper bound $\eta(\omega_{max})^\star$
(given by Eq.~(\ref{bound2})), of  $\eta(\omega_{max})$ is plotted as solid curve. Dotted dashed line corresponds to the Curzon-Ahlborn limit $\eta(\omega_{max})=\eta_{C}/2$.  Note that the Curzon-Ahlborn limit is exceeded in the interval [1,2] with the transmission model. } \label{trans1}
\end{figure}

Compared to the Aharonov-Bohm interferometer model, the rate of variation of maximum efficiency with asymmetry is much less for the transmission model.  Also, the values of the optimized efficiency with large asymmetry are greater than in the Aharonov-Bohm interferometer model by more than one order of magnitude. For instance, when the asymmetry parameter $x=10$, the optimized value of $\eta_{max}=0.0045\eta_{C}$ for the Aharonov-Bohm interferometer model and $0.28\eta_{C}$ for the transmission model. To understand the large difference of efficiency obtained between the two models, we take a closer look on the optimized values of the parameters. Our analysis shows that optimal values of $\eta_{max}$ are obtained when the transmission probabilities $\left(c_{ij}\right)_k$, which in principle could take any value between 0 and 1, are either 0 or 1. In particular, we find that the efficiency is maximal for particular combinations of the transmission probabilities, as described below. Out of the three combinations (one for each window), two correspond to symmetric transmission  ($\left(c_{ij}\right)_k=\left(c_{ji}\right)_k$ for $i,j=L,R,P$) and one to perfectly asymmetric transmission ($\left(c_{ij}\right)_k=1$ and $\left(c_{ji}\right)_k=0$ for some values of $i$ and $j$). Also, for the symmetric cases only two reservoirs are involved in the transmission.  For instance, the optimized values of parameters for $x=1.4$ are plotted in Fig. \ref{opt}. Panel (a) corresponds to the derivative of Fermi distribution function $\frac{\partial f(E)}{\partial E}$  with the optimized temperature $T=1$ and chemical potential $\mu=1.333$. Three energy windows of transmission are 1) $[-3.861,-3.713]$, 2) $[-1.326,-1.325]$, and 3) $[0.4931,0.5201]$. For the first case, $\tau_{LR}=\tau_{RL}=0$, $\tau_{LP}=\tau_{PL}=0$, and $\tau_{RP}=\tau_{PR}=1$. Here, the transmission is perfectly symmetric and is only between the right reservoir  and the probe  (i.e., $R \leftrightarrows P$). Transmission is perfectly asymmetric in the second energy window. This is clear from the figure as $\tau_{LR}=0, \tau_{RL}=1$, $\tau_{LP}=1, \tau_{PL}=0$, and $\tau_{RP}=0,\tau_{PR}=1$. Transport of electrons between the reservoirs is from $L \rightarrow R \rightarrow P \rightarrow L$.  For the third energy window, the transmission probabilities are $\tau_{LR}=\tau_{RL}=0$, $\tau_{LP}=\tau_{PL}=1$, and $\tau_{RP}=\tau_{PR}=0$ and are completely symmetric  between the left reservoir and the probe  (i.e., $L \leftrightarrows P$). We also obtained similar results using models with three delta peaks for transmission instead of transmission windows. Note that for the delta peaks model the width of transmission windows tends to zero i.e., $\Delta_{k}\rightarrow0$.

Curzon-Ahlborn limit is slightly exceeded for small values of asymmetry in the Aharonov-Bohm interferometer model. Can the efficiency at maximum power go significantly beyond the Curzon-Ahlborn limit? To check this we optimize efficiency at maximum power $\eta(\omega_{max})$. Results of our optimization are shown in Fig. \ref{trans1}. As expected, Curzon-Ahlborn limit $\eta(\omega_{max})=\eta_{C}/2$ is exceeded. When $x>0$, $\eta(\omega_{max})$ increases quadratically with increase in the asymmetry parameter until $x=1.349$. At $x=1.349$, it takes the maximum value of $0.569\eta_{C}$. Beyond $1.349$, $\eta(\omega_{max})$ decreases. Moreover, for the asymmetry parameter $x$ in the interval $[1,2]$, $\eta(\omega_{max})>\eta_{C}/2$.  Our numerical analysis shows that the optimized values of transmission probabilities corresponding to the above interval is completely symmetric for two energy windows and completely asymmetric for the third window as discussed in the Fig. \ref{opt}. Also, outside the interval the optimized values of the probabilities are completely symmetric for two windows whereas for the third window it is not completely asymmetric between the different reservoirs. It is clear from the definition of transmission probabilities in Eq. (\ref{tp}) that Aharonov-Bohm interferometer model with three quantum dots discussed in the previous section cannot realize the above mentioned transmission probabilities that leads to $\eta(\omega_{max})$ largely above the Curzon-Ahlborn limit. A more complicated Hamiltonian with large number of levels might approach a one to one correspondence between the results in the transmission and Aharonov-Bohm interferometer model.

 Similarly to  maximum efficiency, at large values of $|x|$, the efficiency at maximum power decreases with $|x|$. Also, for $x<0$, $\eta(\omega_{max})$ increases initially before the decrease (with the maximum value $\eta(\omega_{max})=0.24261$ obtained at $x=-50$).
Note that for large values of asymmetry both the maximum efficiency and
the efficiency at the maximum power are much smaller than the
upper bounds set by thermodynamics (as discussed in Ref.~\cite{prlbts})
and shown as solid curves in the Figs. \ref{trans} and \ref{trans1}.

We have also increased the number of transmission windows, up to
$n=7$ (data not shown), and found only a slight increase of the maximum value
of $\eta(\omega_{\rm max})$ up to $0.5709\eta_C$ at $n=7$.

Our numerical results suggest that a three terminal thermoelectric transport yields large asymmetry with very low efficiency. Does the thermodynamics impose bound on maximum value of efficiency in non-interacting systems with broken time reversal symmetry under three terminal transport? Is it possible to achieve Carnot efficiency with asymmetry in the system under three terminal transport? These questions are addressed in the next subsection by considering a general model with random values for the Onsager coefficients.

\subsection{Random Onsager matrix model}

Consider then a $4\times4$ random Onsager matrix $\mathbf{L}$. The elements $L_{ij}$ are  chosen from the uniform distribution $[-1,1]$ and are directly substituted in Eq. (\ref{flux2}) to calculate the reduced Onsager matrix $\mathbf{L}'$. In general, the elements of the Onsager matrix are not independent and are bound to fulfill certain conditions. First of all, it follows from thermodynamics that the entropy production rate has to be positive i.e., $\dot{S}=J_{1}X_{1}+J_{2}X_{2}+J_{3}X_{3}+J_{4}X_{4}\geq0$.  This requires that the Onsager matrix $\mathbf{L}$ is positive-definite.

Now, we turn our attention to specific case of non-interacting
three-terminal system. In this case, there are further restrictions on the elements of Onsager matrix $\mathbf{L}$. For instance, it is clear from the Eqs. (\ref{onsager}) and (\ref{onsager1}) that
$L_{13},L_{24},L_{31},L_{42}\leq 0$
(and that, as expected in general, all diagonal elements
$L_{ii}\geq 0$, $i=1,...,4$).
Also, the off diagonal elements of each block matrix are even functions of magnetic field  and hence are equal, i.e., $L_{12}=L_{21}$, $L_{14}=L_{23}$,
$L_{32}=L_{41}$, $L_{34}=L_{43}$. Note that this symmetry of the off diagonal elements is broken for the reduced Onsager matrix $\mathbf{L}'$, where
we can have $L_{12}^\prime \neq L_{21}^\prime$.
Positivity of the transmission probabilities, $\tau_{ji}(E)\geq 0$
and of $-\partial f/\partial E$,
and the sum rule $\sum_{i}\tau_{ij}(E)=\sum_{j}\tau_{ij}(E)$ imply
$L_{11}\geq |L_{13}|,|L_{31}|$;
$L_{22}\geq |L_{24}|,|L_{42}|$;
$L_{33}\geq |L_{13}|,|L_{31}|$; and
$L_{44}\geq |L_{24}|,|L_{42}|$.

 There are further constraints on the values of Onsager matrix elements from the structure of reservoirs, which we assume to be ideal Fermi gases at temperature $T$ and chemical potential $\mu$.
Upper bound on the energy integrals appearing in the Onsager coefficients are
then given by
  \begin{eqnarray}
  \frac{e^2 T}{h}\int dE \,\tau_{ij}(E)
\left[-\frac{\partial f(E)}{\partial E}\right] &\leq & \frac{T}{2}, \nonumber\\
 \frac{e T}{h}\int dE \, (E-\mu) \tau_{ij}(E)
\left|\left[-\frac{\partial f(E)}{\partial E}\right]\right| &\leq&\frac{T^{2}\ln 2}{2} , \nonumber\\
  \frac{T}{h}\int dE \, (E-\mu)^{2} \tau_{ij}(E)
\left[-\frac{\partial f(E)}{\partial E}\right]&\leq &\frac{T^{3} \pi ^{2}}{24}.
  \end{eqnarray}
The first and the third bound are saturated by setting
$\tau_{ij}(E)=1$ for all values of energy, the second one by setting
$\tau_{ij}=0$ for $E<\mu$ and $\tau_{ij}=1$ otherwise,
or vice versa $\tau_{ij}=0$ for $E>\mu$ and $\tau_{ij}=0$ otherwise.

Even with all these restrictions, we find that a random Onsager matrix model can saturate the upper bounds $\eta_{\max}^\star$ and $\eta(\omega_{max})^\star$
from thermodynamics for efficiencies
(solid curves in Figs. \ref{trans} and \ref{trans1}).
In particular,
$\eta(\omega_{max})\rightarrow \eta_{C}$
when $|x| \rightarrow \infty$.
However, it is clear from the definition of Onsager coefficients in Eqs. (\ref{onsager}) and (\ref{onsager1}) that for a non-interacting system there are further correlations between the different Onsager coefficients. For instance, the different integrands are weighed by $(E-\mu)^{n}, n=0,1,2$.  For small systems, in particular for a three level system, these correlations are significant and put bounds on maximum achievable efficiency as implied by our optimization results.

\section{Conclusions}

We have investigated the thermoelectric efficiency of broken time reversal symmetry systems under three terminal transport. In these systems, the efficiency is determined by two parameters, the asymmetry parameter $x$ and the ``figure of merit''
$y$. First, we have studied the optimized efficiency of a realistic model of Aharonov-Bohm interferometer formed with three non-interacting quantum dots using simulated annealing.  Our results show that Carnot efficiency $\eta_{C}$ can be obtained when the thermopower is symmetric ($x=1$). Introducing asymmetry in thermopower, the maximum efficiency $\eta_{max}$  decreases from $\eta_{C}$. However, the efficiency at maximum power $\eta(\omega_{max})$  is maximal with asymmetry in thermopower ($x\neq1$). In particular, our studies illustrate that Curzon-Ahlborn limit can be exceeded in the linear response regime for a realistic model with broken time reversal symmetry. We note that our results could be of experimental relevance in view of the recent progress in the phase-coherent manipulation of heat in solid-state nanocircuits, see \cite{expt1,dubi,expt2} and references therein.
We have also studied the thermoelectric efficiency of a generic model using random Hamiltonians drawn from GUE. Our analysis shows that it is highly improbable to obtain large values of both asymmetry in thermopower and efficiency.

Furthermore, we have shown that the efficiency can be improved using an energy dependent transmission. In particular, optimizing the transmission matrix elements at three energy windows we have found more than one order of magnitude increase in the efficiency at large asymmetry over the three level Aharonov-Bohm interferometer model. The optimal values of transmission probabilities at each energy window are either $0$ or $1$. Also, two energy windows correspond to symmetric transmission and one to perfectly asymmetric transmission. The Curzon-Ahlborn limit is exceeded in the interval $[1,2]$ of the asymmetry parameter $x$ and $\eta(\omega_{max})$ as large as $0.57\eta_{C}$ is obtained. Similarly to the Hamiltonian models considered in this paper, the efficiency decreases at large values of asymmetry.

Our extensive and accurate numerical results suggest that a three terminal thermoelectric transport is viable only for large asymmetry with very low efficiency. On the other hand, using a model with random values of Onsager coefficients one may obtain Carnot efficiency for maximum efficiency $\eta_{max}$ and efficiency at maximum power  $\eta(\omega_{max})$ for arbitrary large values of asymmetry. However, we argue that for non-interacting systems with a small number of levels there are correlations between the Onsager coefficients that bound the efficiency. In order to obtain large efficiency for large asymmetry, we need to turn our attention to systems with more than three terminals or to interacting systems or to go beyond the linear response.
This remains to be analyzed in future.

\emph{Note.} After completion of our work, we became aware of a related
work~\cite{saitonew}, showing the existence of upper bounds on thermodynamic
efficiencies for three-terminal transport as a consequence of the unitarity
of the scattering matrix. The results from our transmission model,
depicted in Figs.~\ref{trans} and \ref{trans1}, in practice saturate these new
bounds.


\begin{thebibliography}{100}

  \bibitem{te} G. Mahan, B. Sales, and J. Sharp, Phys. Today {\bf 50}, 42 (1997).

    \bibitem{Majumdar} A. Majumdar, Science {\bf 303}, 777 (2004).

\bibitem{dresselhaus}
M. S. Dresselhaus, G. Chen, M. Y. Tang, R. G. Yang, H. Lee,
D. Z. Wang, Z. F. Ren, J. -P. Fleurial, and P. Gogna,
Adv. Mater. \textbf{19}, 1043 (2007).

\bibitem{snyder}
G. J. Snyder and E. S. Toberer,
Nature Mater. \textbf{7}, 105 (2008).

\bibitem{shakuori}
A. Shakouri,
Annu. Rev. Mater. Res. {\bf 41}, 399 (2011).

\bibitem{dubi}
Y. Dubi and M. Di Ventra,
Rev. Mod. Phys. {\bf 83}, 131 (2011).

\bibitem{BC11}
G. Benenti and G. Casati,
Phil. Trans. R. Soc. A {\bf 369}, 466 (2011).
\bibitem{prlbts} G. Benenti, K. Saito, and G. Casati,
Phys. Rev. Lett. {\bf 106}, 230602 (2011).
\bibitem{vandenbroeck}
C. Van den Broeck,
Phys. Rev. Lett. {\bf 95}, 190602 (2005).
   \bibitem{curzon1} J. Yvon, {\it Proceedings of the International Conference on Peaceful Uses of Atomic Energy} (United Nations, Geneva, 1955), p. 387.
     \bibitem{curzon2} P. Chambadal, {\it Les Centrales Nucl\'{e}aires} (Armand Colin, Paris, 1957).
        \bibitem{curzon3} I. I. Novikov, J. Nucl. Energy {\bf 7}, 125 (1958).
     \bibitem{curzon4} F. Curzon and B. Ahlborn, Am. J. Phys, {\bf 43}, 22 (1975).
\bibitem{esposito2009}
M. Esposito, K. Lindenberg, and C. Van den Broeck,
Phys. Rev. Lett. {\bf 102}, 130602 (2009).

\bibitem{schulman}
B. Gaveau, M. Moreau, and L.S. Schulman,
Phys. Rev. Lett. {\bf 105}, 060601 (2010).

\bibitem{esposito2010}
M. Esposito, R. Kawai, K. Lindenberg, and C. Van den Broeck,
Phys. Rev. Lett. {\bf 105}, 150603 (2010).

\bibitem{linke}
N. Nakpathomkun, H. Q. Xu, and H. Linke, Phys. Rev. B. {\bf 82}, 235428 (2010).

\bibitem{seifert}
U. Seifert,
Phys. Rev. Lett. {\bf 106}, 020601 (2011).

\bibitem{goupil}
Y. Apertet, H. Ouerdane, C. Goupil, and Ph. Lecoeur,
Phys. Rev. E {\bf 85}, 031116,
\emph{ibid.}, 041144 (2012).

    \bibitem{astp1}  K. Saito, G. Benenti,  G. Casati, and T. Prosen,
Phys. Rev. B. {\bf 84}, 201306(R) (2011).

\bibitem{astp2} D. S\'anchez and L. Serra, Phys. Rev. B. {\bf 84}, 201307(R) (2011).

\bibitem{imry1}
O. Entin-Wohlman, Y. Imry, and A. Aharony,
Phys. Rev. B {\bf 82}, 115314 (2010).
\bibitem{buttiker}
R. S\'anchez and M. B\"uttiker,
Phys. Rev. B {\bf 83}, 085428 (2011).
\bibitem{imry2}
J.-H. Jiang, O. Entin-Wohlman, and Y. Imry,
Phys. Rev. B {\bf 85}, 075412 (2012).
\bibitem{ora}
O. Entin-Wohlman and A. Aharony, Phys. Rev. B. {\bf 85}, 085401 (2012).
\bibitem{buttiker2}
B. Sothmann, R. S\'anchez, A. N. Jordan, and M. B\"uttiker
Phys. Rev. B {\bf 85}, 205301 (2012).

\bibitem{railway} M. Horvat, T. Prosen, G. Benenti, and G. Casati,
Phys. Rev. E {\bf 86}, 052102 (2012).
\bibitem{sti} W. H. Press, B. P. Flannery, S. A. Teukolsky, and W. T. Vetterling, {\it Numerical Recipes in Fortran 77: The Art of Scientific Computing} (Cambridge University Press, Cambridge, 1992).
 \bibitem{Landauer} S. Datta, {\it Electronic Transport in Mesoscopic Systems} (Cambridge University Press, Cambridge, 1995).
\bibitem{footnote:cost}
In the optimization, $\eta_{max}$ at a particular value of $x_c$ of the asymmetry parameter is obtained by subtracting a cost function $|x-x_{c}|$ from the optimizing function $\eta_{max}$.
\bibitem{haake}
See, for instance,
F. Haake, {\it Quantum Signatures of Chaos}, 2nd. ed.
(Springer-Verlag, Berlin, 2000).
\bibitem{expt1}
F. Giazotto, T. T. Heikkil\"a, A. Luukanen, A. M. Savin, and J. P. Pekola,
Rev. Mod. Phys. \textbf{78}, 217 (2006).
\bibitem{expt2} F. Giazotto and M. J. Mart\'inez-P\'erez, Nature {\bf 492}, 401 (2012).
\bibitem{saitonew} K. Brandner, K. Saito, and U. Seifert,
Phys. Rev. Lett. {\bf 110}, 070603 (2013).

        \end{thebibliography}
  \end{document}